\documentclass[preprint,showpacs,preprintnumbers,pra,floatfix,superscriptaddress]{revtex4}

\usepackage[T1]{fontenc}
\usepackage[pdftex]{graphicx,color}
\usepackage{bm}        
\usepackage{amssymb}   
\usepackage{amsmath}
\usepackage[T1]{fontenc} 
\usepackage{threeparttable}
\usepackage{cancel}
\begin{document}

\title{Reexamination of the ground state Born-Oppenheimer Yb$_2$ potential}

\author{Giorgio Visentin}\email{giorgio.visentin@skoltech.ru}
\affiliation{Skolkovo Institute of Science and Technology, Skolkovo Innovation Center, Moscow  121205,  Russia}

\author{Alexei A. Buchachenko}\email{a.buchachenko@skoltech.ru}
\affiliation{Skolkovo Institute of Science and Technology, Skolkovo Innovation Center, Moscow  121205,  Russia}
\affiliation{Institute of Problems of Chemical Physics RAS, Chernogolovka, Moscow Region 142432, Russia}

\author{Pawe{\l} Tecmer}\email{ptecmer@fizyka.umk.pl}
\affiliation{Institute of Physics, Faculty of Physics, Astronomy and Informatics, Nicolaus Copernicus University in Toru\'n, Grudziadzka 5, 87-100 Torun, Poland}

\begin{abstract}
The precision of the photoassociation spectroscopy of Yb dimer in degenerate gases is enough to improve the constraints on the new short-range gravity-like forces if the theoretical knowledge of the Born-Oppenheimer interatomic potential and non-Born-Oppenheimer interactions is refined [M. Borkowski et al. Sci. Rep. A {\bf 9}, 14807 (2019)]. The ground-state interaction potential of ytterbium dimer is investigated at the eXact 2-component core-correlated CCSD(T) level of {\it ab initio} theory in the complete basis set limit with extensive augmentation by diffuse functions. For the small basis set the comparison is made with the four-component relativistic finite-nuclei CCSD(T) calculations to identify the contraction of the dimer bond length as the main unrecoverable consequence of the scalar-relativistic approximation. Empirical constraint on the number of bound vibrational energy levels of the $^{174}$Yb$_2$ dimer is accounted for by representing the global {\it ab initio}-based Born-Oppenheimer potential with the model semianalytical function containing the scale and shift parameters. The results support the previous evaluation of the Yb dimer potentials from the  photoassociation spectroscopy data and provide an accurate and flexible reference for future refinement of the constraints on the short-range gravity-like forces by ultracold atomic spectroscopy. 
\end{abstract}

\maketitle


\section{Introduction}
\label{Sec:Intro}

Recent advances in ultracold atomic physics have brought weakly-bound ytterbium dimer, never detected spectroscopically at room temperature, up to forefront fundamental and applied research. Atomic energy level scheme convenient for laser confinement, cooling, and narrow band excitation was explored in the ultraprecise Yb frequency standards for time measurements \cite{Ludlow2015}  and relativistic geodesy applications \cite{Ludlow2018}. Yet, further refinement has been predicted for a clock based on Yb dimer \cite{Borkowski2018}. Numerous photoassociation spectroscopy (PAS) studies in the degenerate Yb gases made significant insight in ultracold collision dynamics, long-range interactions and exotic diatomic states \cite{Tojo2006, Enomoto2008, Borkowski2009, Takasu2012, Kato2013, Green2016, Takasu2017, Borkowski2017, Franchi2017, Bouganne2017, Cappellini2019}. A variety of naturally abundant isotopes of bosonic and fermionic nature made Yb dimer very attractive for studying spectroscopic manifestations of the non-classical mass-dependent effects \cite{Borkowski2017, Lutz2016}. 

Particularly challenging is the use of PAS measurements to constrain the short-range gravity-like forces at the nanometer scale to affirm and improve the results of the neutron scattering and atomic force microscopy experiments \cite{Borkowski2019}. It has been shown that unprecedented 0.5 kHz accuracy of the PAS data on the near-threshold rovibrational levels of Yb$_2$ is enough to improve existing constraints on the Yukawa-type forces \cite{Fayet1996,Adelberger2009} by almost two orders of magnitude. This, however, requires essential refinement of the underlying theoretical model for PAS spectroscopy. 

The BBO (beyond-Born-Oppenheimer) model introduced in Ref.~\cite{Borkowski2017} relied on the Born-Oppenheimer (BO) reference, in which the {\it ab initio} potential for the Yb$_2$ ground $^1\Sigma^+_g (0^+_g)$ state \cite{Buchachenko2007} was used with the adjustable well-depth scaling parameter and the long-range tail. The BBO terms included were the adiabatic mass-dependent corrections \cite{Lutz2016,Pachucki2008} and isotopic field shift  \cite{Lutz2016}. Initial approximations to each term were taken from different sources and also contained adjustable parameters. The best fits to the experimentally measured energies of 13 near-threshold levels of the bosonic isotopomers $^{168,170,174}$Yb$_2$ gave the root-mean-square (RMS) deviations in the range of 113 kHz (3.8$\times$10$^{-6}$ cm$^{-1}$) and 30 kHz (1.0$\times$10$^{-6}$ cm$^{-1}$) for the reference BO and BBO models, respectively.

Despite this seemingly impressive agreement, theoretical RMS deviations are 60 times larger than the experimental uncertainty. Taking into account a limited number of the levels probed experimentally and their predominant sensitivity to the long-range interatomic interactions, a plausible way of refining the theory is to obtain more robust initial guesses for molecular parameters though the more thorough elaboration of the non-adiabatic corrections and inclusion of subtle relativistic and quantum electrodynamic effects \cite{Borkowski2019} by no means can be discarded. An obvious first step is the refinement of the BO Yb$_2$ potential, which contributes by far major share of uncertainty and is not defined globally with due veracity \cite{Borkowski2009,Kitagawa2008,Buchachenko2007,WangDolg,Mosyagin2010,Tecmer2019}.

Indeed, this potential is accurately defined only in the long-range region. The  dispersion coefficients  were the subject of reliable {\it ab initio} calculations \cite{Zhang2008, Safronova2012, Porsev2014, Visentin2019}. Besides, very high sensitivity of the ultracold data to the lowest-order $C_6$ and $C_8$ coefficients permitted their high-precision adjustment \cite{Borkowski2009, Borkowski2017, Kitagawa2008}. However, PAS and scattering length data are not sensitive enough to interatomic interaction at short and medium ranges. In terms of the semiclassical analysis by Gribakin and Flambaum \cite{GF1993}, they can fix the area of the potential minimum, but not its position, $R_e$, and depth, $D_e$. The number of bound vibrational energy levels $N_\mathrm{max}$ is therefore known and $N_\mathrm{max} = 72$ for the reference $^{174}$Yb$_2$ dimer was used for testing the BO models \cite{Borkowski2009, Borkowski2017, Kitagawa2008}. Model potentials satisfying this condition with $R_e = 4.50$ {\AA} and $D_e = 1083$ cm$^{-1}$ were able to fit the PAS data with the RMS deviation of 54 kHz (1.8$\times$10$^{-6}$ cm$^{-1}$) \cite{Borkowski2009, Kitagawa2008}, better than the BO model did with the {\it ab initio} potential having $R_e = 4.55$ {\AA} and $D_e$ adjusted to 743 cm$^{-1}$ (113 kHz cited above) \cite{Borkowski2017}. Worse still, no room-temperature experimental data attesting Yb$_2$ potential minimum do exist. An exception is mass-spectrometric dissociation energy estimation of  1400 $\pm$ 1400 cm$^{-1}$~\cite{Guido1972}. To that end, it is extremely important to provide a reliable {\it ab initio} electronic structure model for Yb$_2$ and fully understand its limitations.

Consistent {\it ab initio} treatment of all the effects physically important for the interaction of two Yb atoms is not feasible. Predominantly dispersion bonding character implies extensive recovery of the dynamic electron correlation, core-valence and outer-core one included, in a highly saturated basis set heavily augmented with a diffuse component. The scalar-relativistic (SR) coupled cluster method with singles, doubles, and non-iterative triples, CCSD(T), and extrapolation to the complete basis set (CBS) limit provides the only tractable means to meet these requirements. It disregards, however, essential contributions from high-order cluster excitations. Furthermore, vectorial relativistic effects, such as spin-orbit coupling (SOC), are not negligible for lanthanides \cite{Pekka1988}. To estimate them quantitatively, one should assess the SR results against the sophisticated four-component calculations, far more restrictive with regards to the basis set size and extent of the correlation treatment. All together led to a considerable mismatch between the literature {\it ab initio} results for Yb$_2$~\cite{Buchachenko2007, WangDolg, Mosyagin2010, Tecmer2019}, which calls into question what should be considered as a reference. The first goal of the present work is therefore to address that challenge by performing, up to date,  the most reliable all-electron SR CCSD(T) calculations and assessing their convergence. Being performed within the conventional {\it ab initio} frames, our analysis is expected to be useful for similar dispersion-bound molecules containing lanthanide atoms. The second goal is to provide improved representation of the global BO Yb$_2$ potential with its preliminary uncertainty estimation. We thus describe the semiclassical scaling of the CCSD(T) potentials to make a preliminary assessment of their compliance with the ultracold data, explore the sensitivity to the parameters of the potential minimum and estimate the BO contributions not included in the {\it ab initio} calculations.  On the one hand, our results support the previous BO and BBO models \cite{Borkowski2017, Borkowski2019}, on the other -- provide more reliable and flexible reference BO potential function for use in the more sophisticated BO or BBO spectroscopic models. 

The next Section combines detailed descriptions of the {\it ab initio} approaches and results. The high degree of technicality is indispensable there to make sure that our findings are reproducible and transferable. In Section ~\ref{sec:model}, we present the results of the semiclassical analysis of the {\it ab initio}-based global BO potential and discuss its implications to the BO and BBO modeling. Conclusions follow. 

\section{{\it Ab initio} CCSD(T) calculations}
\label{sec:AbinitNew}

\subsection{Scalar-relativistic calculations with \textsc{Molpro}}
\label{sec:abiscalar}

The CCSD(T) calculations were performed using the exact two-component (X2C) scalar-relativistic Hamiltonian
\cite{Ricatti-proceedings,Ricatti,X2C,Sikkema-X2C,Liu-X2C,Peng2012} as realized in the \textsc{Molpro} 2015.1 program package \cite{MOLPRO}. In all our calculations the energy convergence threshold was set to 10$^{-10}$ $E_h$. The sequence of the correlation-consistent polarized valence cc-pV$n$Z basis sets with the cardinal numbers $n$ = D, T, Q (hereinafter for brevity V$n$Z) contracted for use with the X2C approximation \cite{YbAEBas} was used. 
To compensate for the lack of optimized diffuse augmentation in the cc-pV$n$Z sets, we added one or two primitives for each symmetry type with the exponents continuing the sequence of basis exponents in an even-tempered manner with the default parameters of the \textsc{Molpro} package. In what follows, these options are denoted as e1 or e2, respectively. The 3s3p2d2f1g set of bond function (bf) \cite{BF} placed at the midpoint of Yb--Yb distance was also used for the same purpose. The calculations were performed in the $D_{2h}$ symmetry group for the dimer and the $C_{2v}$ group for the Yb atom in the full dimer basis to implement the counterpoise (CP) correction \cite{BoysBernardi} using the restricted Hartree--Fock reference functions. Two series of the CCSD(T) calculations correlated $5s$, $5p$, $4f$, and $6s$ orbitals (c46 option, 46 electrons within the core per atom) and, in addition, $4s$, $4p$, and $4d$ orbitals (c28 option). Negligible effect of correlating deeper shells (orbitals below 4s) on the potential energy well parameters and dispersion coefficients of the dimer was proven in Refs.~\citenum{Tecmer2019} and \citenum{Visentin2019}. We should also note in this regard that we encountered a convergence difficulty while using the correlation-consistent polarized weighted core-valence basis sets (cc-pwCV$n$Z) with $n > 2$. Thus, the results with that basis set family, more appropriate for the c28 correlation option, are not reported. The counterpoised potentials obtained with the VDZ, VTZ and VQZ basis sets were extrapolated to the CBS limit using the mixed exponential-Gaussian formula \cite{CBS11, CBS12}. 

A non-uniform grid of 57 internuclear distances spanning the range from 2 to 50 {\AA} was used. At distances longer than 25 {\AA}, erratic non-smooth variations of energies were detected. This prevents a firm determination of the dominant $C_6$ dispersion coefficient by fitting and limits the accuracy of interaction energies by 0.05 cm$^{-1}$.  The Supplemental Materials to this paper tabulate the potential energies obtained with each V$n$Z basis set and extrapolated  to the CBS limit. 

To establish the connection to the relativistic calculations with the \textsc{Dirac} code described in the next subsection, a few auxiliary X2C CCSD(T) calculations were performed on a slightly shorter grid. They used the c48 correlation option ($5p$, $4f$ and $6s$ orbitals correlated) and the original VDZ, uncontracted VDZ (uVDZ), and uncontracted Dyall double-$\zeta$ \cite{dyall_b} bases, all without further augmentation.

\subsection{Assessment of the scalar-relativistic approximation with \textsc{Dirac}}
\label{sec:abiso}

All the CCSD(T) calculations were carried out using the~\textsc{DIRAC19} relativistic software package~\cite{dirac_paper} and utilized a linear $D_{\infty h}^*$ symmetry, the valence double-$\zeta$ basis set of Dyall~\cite{dyall_b}, and the Gaussian nuclear model, if not stated otherwise. The calculations kept $4f$, $5p$, and $6s$ spinors (orbitals) correlated (leaving 48 inner electrons within the core per atom, c48 option) and all virtuals were active.

We assessed the performance of various relativistic Hamiltonians. Specifically, these include (i) the four-component Dirac--Coulomb Hamiltonian (with a default setup~\cite{lvcorr}) denoted as 4C-DC, (ii) the spin-free Dirac--Coulomb Hamiltonian denoted as SF-DC~\cite{dyall-spin-free}, and (iii) the spin-free X2C Hamiltonian \cite{Ricatti-proceedings,Ricatti,X2C,Sikkema-X2C,Liu-X2C,Peng2012} denoted as SF-X2C. An additional set of calculations was performed for the SF-X2C Hamiltonian with the point-charge nuclear model. That allowed us for direct comparison with the SR calculations carried out in the \textsc{Molpro} software package  described above.

As we aimed to analyze the relativistic effects on the equilibrium parameters, the grid of 27 internuclear distances $R$ were restricted to 5--20 a$_0$ (roughly 2.6--10.6 {\AA}) interval. The CP correction \cite{BoysBernardi} was applied imposing the $C_{\infty v}^*$ point group symmetry. Tabulated \textit{ab initio} potential energies with different relativistic Hamiltonians are available in the Supplemental Materials. 

Table~\ref{tab:vdz} compares the parameters of the CCSD(T) c48 Yb$_2$ interaction potentials, namely,  inflection point at zero kinetic energy $\sigma$ and equilibrium parameters $R_e$, $D_e$, as obtained for the basis sets of double-$\zeta$ quality without further augmentations. Implementations of the  X2C Hamiltonian in \textsc{Dirac} and \textsc{Molpro} packages gave identical results for the Dyall basis set. The \textsc{Molpro} calculations with the cc-pV$n$Z-type basis showed a weaker bonding. This indicates that the VDZ basis even in its uncontracted form is not optimal for predominantly dispersion interaction we are dealing with here. The contraction alters $D_e$ marginally, but reduces $R_e$ by almost 0.02 {\AA}. Such a significant increment should persist for the V$n$Z sets of higher cardinal numbers and thus cannot be recovered by extrapolation to the CBS limit. The calculations with \textsc{Dirac} package demonstrated that a slight increase of the binding energy due to SO coupling and finite-nuclei effects is nearly canceled for the SF X2C Hamiltonian. By contrast, related contraction of $R_e$ by 0.01 {\AA} reflects the main deficiency of the scalar-relativistic approximation. 

\begin{table}[htb]
\centering
\caption{Parameters of the CCSD(T) c48 potentials calculated with the double-$\zeta$ basis sets. Calculations with \textsc{Dirac} used Gaussian nuclear model unless stated otherwise.}
\label{tab:vdz}

\begin{tabular}{l@{\quad}c@{\quad}c@{\quad}c@{\quad}c@{\quad}}
\hline
\hline
Method & Package & $\sigma$ ({\AA}) & $R_e$ ({\AA}) & $D_e$ (cm$^{-1}$)\\
\hline
\hline
X2C VDZ & \textsc{Molpro} & 4.080 & 4.741 & 475.4 \\
X2C uVDZ & \textsc{Molpro} & 4.062 & 4.727 & 478.0 \\
X2C Dyall  & \textsc{Molpro} & 4.022 & 4.694 & 531.9 \\
SF-X2C Dyall point charge  & \textsc{Dirac} & 4.022 & 4.694 & 531.9 \\
SF-X2C Dyall  & \textsc{Dirac} & 4.018 & 4.692 & 533.0 \\
SF-DC Dyall  & \textsc{Dirac} & 4.024 & 4.694 & 526.0 \\
4C-DC Dyall  & \textsc{Dirac} & 4.012 & 4.686 & 532.3 \\
\hline
\hline
\end{tabular}
\end{table}

\subsection{{Convergence of the scalar-relativistic results}}
\label{sec:res}

To assess the convergence of the Yb$_2$ X2C CCSD(T) calculations with respect to the diffuse basis augmentation and the core correlation treatment it is instructive to analyze the results at the CBS limit, which accounts for  the convergence with respect to cardinal number of the primary basis set. As the CBS extrapolation procedure cannot be rigorously defined, it introduces additional uncertainty \cite{CBS11,CBS12,CBS21,CBS22,Feller1992,Feller1998,Pilar,Feller2011}. Moreover, additional ambiguity might originate from the CBS extrapolation of the basis set sequences augmented with bond functions. This quite technical issue is discussed in the Supplemental Materials to this paper. 

Table~\ref{tab:params} lists the values of interaction potential parameters after the CBS extrapolation and the semiclassical estimates of the number of bound vibrational energy levels $\mathcal{N}$ obtained with the potential model introduced in Sec.~\ref{sec:model} (\textit{vide infra}). 

The difference between c46 and c28 calculations (all else being equal) indicates that the correlation of the $4s4p4d$ shells reduces the bonding of Yb atoms to a minor extent. That is in line with the previous calculations on the Yb$_2$ potential \cite{Tecmer2019}.  Diffuse augmentation is essential for better recovering the dispersion interactions. While involving the atom-centered even-tempered diffuse primitives (e1) is only suboptimal, it still results in good convergence that follows from the marginal effect of including the second primitive set (e2). The addition of the bond functions enhances Yb$_2$ bonding to a larger extent. For this reason, and for smaller CBS uncertainty due to faster convergence with the cardinal basis number, we took the present V$n$Zbf c28 result as the reference to estimate the converged X2C CCSD(T) potential parameters as $R_e  = 4.585 \pm 0.01$ {\AA} and $D_e = 658 \pm 15$ cm$^{-1}$, where the major share of uncertainties belongs to the form of diffuse augmentation, the minor -- to the CBS extrapolation. The deviation between the X2C CCSD(T) calculations with the cc-pV$n$Z-type and Dyall-type basis sets discussed in Sec.~\ref{sec:abiso} creates additional source of uncertainty. To estimate it better, we consider an artificial CBS extrapolation of the SF-X2C CCSD(T) results for the Dyall VDZ basis set with extrapolation coefficients taken from the corresponding X2C CCSD(T) V$n$Z c48 series. It results in $R_e \approx 4.54$ {\AA} and $D_e \approx 670$ cm$^{-1}$, or $\approx 4.53$ {\AA} and $\approx 700$ cm$^{-1}$ if the increments due to diffuse function augmentation are added. As the physical origin of such a significant mismatch is not clear, the latter values can only be taken as indication that the present calculations likely underestimate the bonding of Yb atoms at the CCSD(T) level of theory.

The results of the previous SR CCSD(T) calculations are presented in Table~\ref{tab:params} as well. The only available all-electron calculations from Ref.~\citenum{Tecmer2019} underestimate the present $D_e$ values by ca. 10\% and overestimate $R_e$ by 0.07 {\AA}. Besides the minor difference in the applied scalar-relativistic Hamiltonians (X2C vs. Douglas--Kroll-Hess of the second order), this mismatch is likely due to the lack of the diffuse functions in the atomic natural orbital relativistic semi-core correlation (ANO-RCC) basis set and the absence of the CP correction. Even though the ANO-RCC basis set was designed to correlate Yb orbitals starting from the 5s shell, the correlation of deeper shells did not result in any problems as reported in Ref.~\citenum{Tecmer2019}. The best ANO-RCC estimation falls between the present VTZe1 and VQZ results, despite the fact that the ANO-RCC basis set was used in its uncontracted form. 

The potential from Refs.~\citenum{Buchachenko2007} is based on the small-core effective core potential ECP28MWB \cite{ECP28MWB} combined with the corresponding ANO basis set \cite{ANO} augmented by the specially designed atom-centered diffuse function set \cite{SC1} and the bf set \cite{BF} simultaneously. It implies stronger bonding of Yb atoms than the present series does, namely, the binding energy is larger by 11\%, and the equilibrium distance is shorter by 1.5\%. This difference should be attributed to the uncertainties of the effective core potential description of the inner shells and of the quality of the ANO basis, as is corroborated by the calculations by Wang and Dolg with the same basis but different diffuse augmentation \cite{WangDolg}. Mosyagin and co-workers \cite{Mosyagin2010} investigated the bonding of Yb atoms using the 28-electron generalized relativistic effective core potential (GRECP) and a series of supplementary basis sets. Only four outer $6s$ electrons were correlated in their reference CCSD(T) calculations with the largest basis. Correction to outer core correlation (OC, equivalent to the present c46 core option) was evaluated with a smaller basis. At this level, their result for binding energy almost falls within the error bars we predicted, but the equilibrium  distance is strongly overestimated.

To go beyond the X2C CCSD(T) approximation, one should consider the vectorial relativistic effects and contributions from the higher-order cluster excitation. For the former, we can use the present 4C-DC calculations. Table~\ref{tab:vdz} indicates that for the double-$\zeta$ basis, the main effect with respect to X2C approximation is the contraction of the dimer bond length. Artificial CBS extrapolation with the coefficients derived from the X2C CCSD(T) V$n$Z c48 series infers the shrinkage of $R_e$ by 0.01 {\AA} and increase of $D_e$ by 9 cm$^{-1}$. Ref.~\citenum{Mosyagin2010} reports the correction twice as large (Table~\ref{tab:params}, ``+SOC''), but the quasi-relativistic two-component density functional theory employed therein is certainly much less accurate. Mosyagin et al. \cite{Mosyagin2010} provided the only and indirect estimation for the higher-order cluster corrections. Iterative contributions of triples and quadruples were recovered by subtracting CCSD(T) energy from the full configuration interaction energy, both obtained by correlating four outer electrons in a medium-size basis set, ``+iTQ'' in the Table~\ref{tab:params}. The effect, 1.5\% reduction of $R_e$ and 20\% increase of $D_e$, is quite significant. However, it is not possible to guess how it would behave upon expanding the basis set and the depth of the shells included in the correlation treatment.

Adding the present estimate for SO coupling and the iTQ correction from Ref.~\citenum{Mosyagin2010} to the converged X2C results cited above, we obtain $R_e \approx 4.45$ {\AA} and $D_e \approx 825$ cm$^{-1}$ as a guess for the true Yb$_2$ BO potential. Uncertainties of these values are hard to quantify, but they are likely not less than 0.1 {\AA} and 100 cm$^{-1}$, respectively.

\begin{table}[htb]
\centering
\caption{{Parameters of the present extrapolated X2C CCSD(T) Yb$_2$ potentials and that from literature. }}
\label{tab:params}
\begin{tabular}{lc@{\quad}c@{\quad}c@{\quad}c@{\quad}}
\hline
\hline
Method & $\sigma$ ({\AA}) & $R_e$ ({\AA}) & $D_e$ (cm$^{-1}$) &  $\mathcal{N}$   \\
\hline
\hline
V$n$Z c46 & 3.916  & 4.593  & 617.5  & 66.99 \\
V$n$Ze1 c46	& 3.907 & 4.596 & 646.1  & 68.15  \\
V$n$Ze2 c46	& 3.907 & 4.598 & 646.9  & 68.20 \\
V$n$Zbf c46	& 3.906 & 4.594 & 654.6 & 68.35 \\
V$n$Z c28	& 3.915 & 4.590 & 615.0 & 66.90 \\
V$n$Ze1 c28 	& 3.906 & 4.596 & 643.6  & 68.12 \\
V$n$Zbf c28	& 3.896	& 4.585 & 657.5  & 68.54 \\
ANO-RCC c28 \cite{Tecmer2019} & &  4.665 & 580 & \\		
ECP28MWB \cite{Buchachenko2007} & 3.870 & 4.522 & 723.7 & 70.97	 \\
ECP28MWB \cite{WangDolg} & & 	4.549 & 742 & \\			
28e GRECP+OC \cite{Mosyagin2010} & &	4.683 & 642	& \\	
28e GRECP +OC+iTQ \cite{Mosyagin2010} & & 4.615 & 767	 &  \\		
28e GRECP +OC+iTQ+SOC \cite{Mosyagin2010} & & 4.582 & 787 & \\		
\hline
\hline
\end{tabular}
\end{table}

\section{Global BO potential by semiclassical scaling}
\label{sec:model}

As was already mentioned, PAS and scattering length measurements established the number of the bound vibrational levels, $N_\mathrm{max}$, of Yb$_2$ dimer.  Previous analysis indicates that this condition can be applied to the BO potentials, as the BBO contributions are too small to alter the number of levels \cite{Borkowski2017}. 

To test the present {\it ab initio} results, we accepted the semianalytical representation of the global BO potential $V(R)$ introduced in Ref.~\citenum{Borkowski2017}:
\begin{equation}
V(R)=[1-f(R)]sV_\mathrm{SR}(R)+f(R) V_\mathrm{LR}(R),
\label{eq:Vfun}
\end{equation}
where the {\it ab initio} points interpolated by cubic splines stand for the short-range part $V_\mathrm{SR}$ and the long-range part contains two lowest dispersion interaction terms 
\begin{equation}
V_\mathrm{LR}(R)=-C_6/R^6 -C_8/R^8,
\label{eq:VLR}
\end{equation}
as the PAS data is not sensitive to the next $C_{10}$ term \cite{Borkowski2017}. The switching function has the fixed form 
\begin{equation}
f(R) = \begin{cases} 0, & \mathrm{if } \;  R \le a  \\
\frac{1}{2}+\frac{1}{4}\sin{\frac{\pi x}{2}}\left(3-\sin^2{\frac{\pi x}{2}}\right), & \mathrm{if } \; a < R < b  \\
1 &  \mathrm{if } \; R > b,  
\end{cases} 
\label{eq:ffun}
\end{equation}
with $x=[(R-a)+(R-b)]/(b-a)$, $a = 10$ a$_0$ (5.292 {\AA}), $b = 19$ a$_0$ (10.054 {\AA}). The values of the  dispersion coefficients $C_6 = 1937.27$ and $C_8 = 226517$ a.u. were taken as fitted in Ref.~\citenum{Borkowski2017} and kept fixed. The scaling parameter $s$ is adjustable. If $s=1$, Eq.(\ref{eq:Vfun}) describes the interpolation of the original {\it ab initio} points at $R \le a$. 

The number of the bound vibrational levels was obtained semiclassically following the procedure described in Ref.~\citenum{GF1993}. The semiclassical phase at zero kinetic energy is given by
\begin{equation}
\Phi = \frac{1}{\hbar} \int_\sigma^\infty \sqrt{2\mu [-V(R)]} dR, 
\label{eq:phase}
\end{equation}
where $\mu$ is the reduced mass of the Yb dimer, for which we used the atomic reduced mass of the $^{174}$Yb$_2$ dimer. Note that Eqs.(\ref{eq:Vfun}), (\ref{eq:ffun})  permit analytical integration of the phase from $R = b$ to infinity that greatly facilitates accurate numerical evaluation of the integral (\ref{eq:phase}). Then $N_\mathrm{max}=[\mathcal{N}]$, $\mathcal{N} =\Phi/\pi + 3/8$. The $\mathcal{N}$ values for the original ($s=1$)  X2C CCSD(T)  potentials are given in Table~\ref{tab:params}. The best of them consistently support 68 bound levels. 

\begin{table}[h]
\centering
\caption{Scaling dimensionless factors $s$ and binding energies (cm$^{-1}$) of the X2C CCSD(T) potentials for the $^{174}$Yb$_2$ dimer.}
\label{tab:scale}
\begin{tabular}{l@{\quad}c@{\quad}c@{\quad}c@{\quad}c@{\quad}c@{\quad}}
\hline
\hline
Potential & $s_\mathrm{min}$ & $s_\mathrm{min}D_e$ & $s_\mathrm{max}$ & $s_\mathrm{max}D_e$ & mean $D_e$ \\
\hline
\hline
V$n$Z c46	    & 1.233  & 761.4  & 1.281 & 791.0 & 776$\pm$15 \\
V$n$Ze1 c46 & 1.172  & 757.2  & 1.218 & 786.9 & 772$\pm$15 \\
V$n$Ze2 c46 & 1.170   & 757.2  & 1.215 & 786.0 & 772$\pm$15 \\
V$n$Zbf c46  & 1.162  & 760.6 & 1.208 & 790.7  & 776$\pm$15 \\ 
V$n$Z c28       & 1.238   & 761.4  & 1.286 & 790.9 & 776$\pm$15 \\
V$n$Ze1 c28 & 1.174   & 755.6  & 1.219 & 784.5 & 770$\pm$15 \\
V$n$Zbf c28 & 1.153   & 758.0  & 1.198 & 787.7 & 773$\pm$15 \\
\hline
\hline
\end{tabular}
\end{table}

\begin{figure}[htb]
\includegraphics[width=0.6\linewidth]{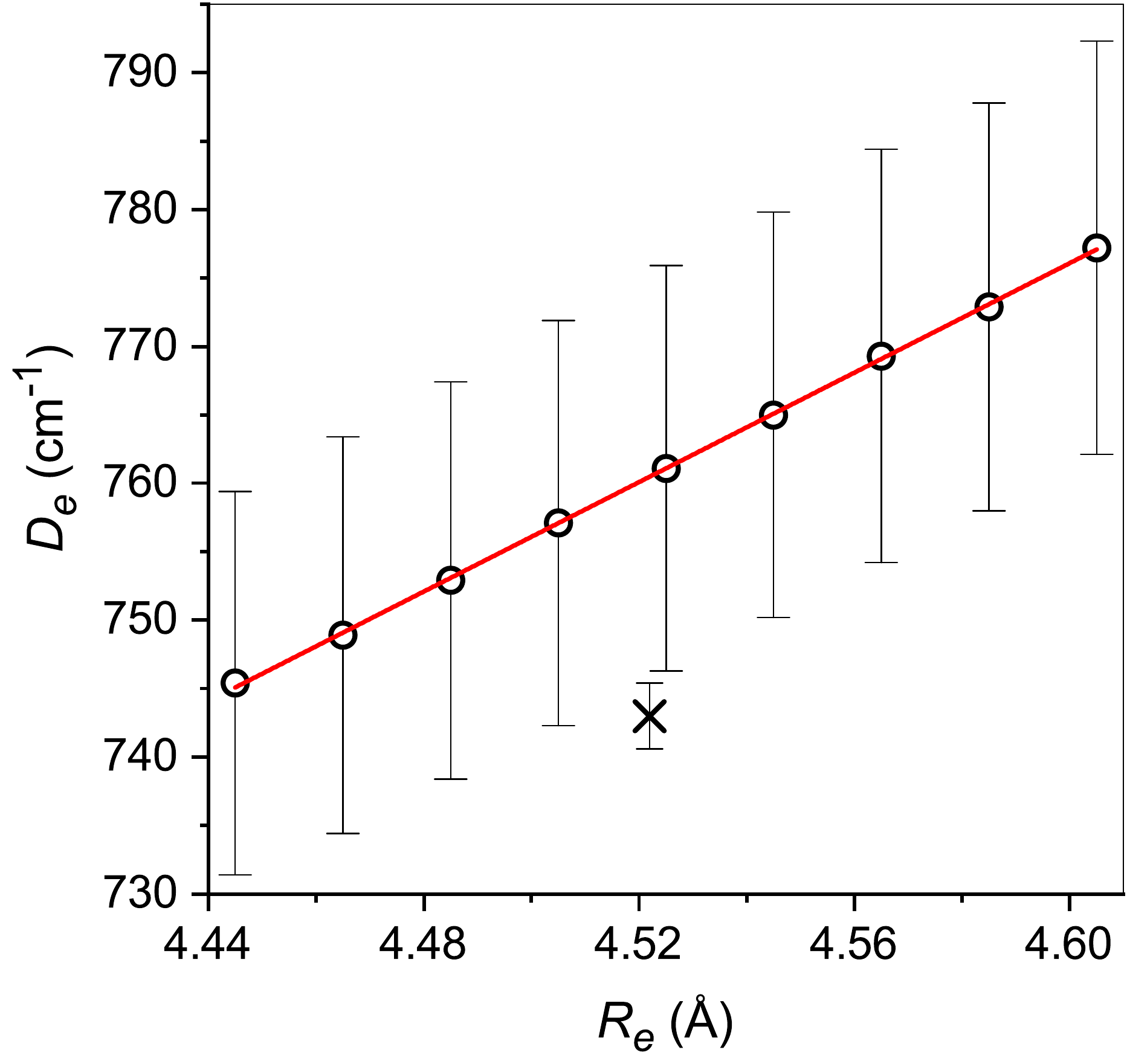}
\caption{Correlation of the scaled binding energy and shifted equilibrium distance of the Yb$_2$ interaction potential based on the {\it ab initio} V$n$Zbf c28 data. Solid lines represents the linear fit}. Cross indicates the scaled Born-Oppenheimer result from Ref.~\citenum{Borkowski2017}.  
\label{fig:scale}
\end{figure}

Next, we vary parameter $s$ until $\mathcal{N}$ exceeds 72 to get $s_\mathrm{min}$ and until $\mathcal{N}$ remains less than 73 to get $s_\mathrm{max}$. The resulting values of $s$, scaled binding energies and mean binding energies computed with the X2C CCSD(T) $V_\mathrm{SR}$ potentials are listed in  Table~\ref{tab:scale}. First, it is evident that the constraint on the bound level projects the original potentials, whose binding energies vary by 43 cm$^{-1}$, to much narrower intervals of 6 cm$^{-1}$ for the limiting and mean values. Such mapping reflects the similarity in the overall shapes of the attractive potentials calculated using the different {\it ab initio} schemes. Second, the range of $D_e$ values that obey the constraint is 30 cm$^{-1}$. As follows from the numerical solutions of the vibrational Schr{\"o}dinger equation, the position of the last near-threshold $N_\mathrm{max}-1 = 71$ level in the same range of $s$ varies by ca. 0.02 cm$^{-1}$. It gives a clue that uncertainty in fixing the near-threshold levels (e.g., by PAS data) gains three orders of magnitude upon propagation to the equilibrium properties.

Preceding analysis of the {\it ab initio} results points out that X2C CCSD(T) CBS calculations significantly overestimate the dimer equilibrium distance. The shift of the potential towards shorter range should obviously affect the scaling by increasing the phase $\Phi$. To take this uncertainty into account, we introduced the shift parameter $d$ and replaced the {\it ab initio} part of the model function (\ref{eq:Vfun}) by the shifted potential $V_\mathrm{SR}(R-d)$. For each $d \in [-0.02,0.14]$ the limiting scaling parameters $s_\mathrm{min}$ and $s_\mathrm{max}$ were found using the same criteria as above. Figure~\ref{fig:scale} shows the results in terms of shifted and scaled equilibrium parameters for the V$n$Zbf c28 {\it ab initio} potential.
Position and the mean depth of the potential well are correlated linearly with the coefficient $200\pm 1$ cm$^{-1}$/{\AA}.  The limits of $D_e$ variation weakly depend on $d$ and amount to $\pm15$ cm$^{-1}$. If we add the uncertainty of the scaled dissociation energies from the Table~\ref{tab:scale}, the margins expand to $\pm18$ cm$^{-1}$.  Figure~\ref{fig:scale} also shows the result of Ref.~\citenum{Borkowski2017}, where the ECP28MWB potential \cite{Buchachenko2007} was used as the reference. It lies on the extended lower bound of the present calculations. We believe that  the Figure~\ref{fig:scale} identifies the most probable ranges for equilibrium parameter variations, i.e., $R_e  \in [4.45, 4.55]$ {\AA} and $D_e = 755 \pm 20$ cm$^{-1}$.

Another implication of the present analysis is the possibility to replace the model function (\ref{eq:Vfun}) with a more flexible form 
 \begin{equation}
V(R)=[1-f(R)]s(d)V_\mathrm{SR}(R-d)+f(R) V_\mathrm{LR}(R),
\label{eq:Vfun1}
\end{equation}
where the constraint $s(d) = -0.304d + 1.176$ results from the correlation between $D_e$ and $R_e$, $V_\mathrm{SR}$ is the present X2C CCSD(T) V$n$Zbf c28 pointwise potential, whereas the Eqs.(\ref{eq:VLR}) and (\ref{eq:ffun}) and their parameters remain unchanged. Keeping a single adjustable parameter ($d$ instead of $s$ in Eq. (\ref{eq:Vfun})), new function features the sensitivity to both position and depth of the potential minimum and thus should have better properties for accurate BO or BBO fitting of the PAS data.

{\section{Conclusions}} 

Present calculations set the scalar-relativistic CCSD(T) benchmark for the ground-state potential of Yb dimer. Achieving the convergence with respect to the basis set saturation and the extent of core correlation treatment, we estimated its equilibrium parameters as {$R_e  = 4.585 \pm 0.01$ {\AA}, $D_e = 658 \pm 15$ cm$^{-1}$ and explained the main sources of disagreements in the previous {\it ab initio} calculations. The first four-component relativistic CCSD(T) calculations performed with a double-$\zeta$ basis generally support the validity of the X2C approximation but pointed to the bond length shrinkage of about 0.01 {\AA} as the main effect caused by relativistic contraction. The quoted uncertainties, not exceeding 3\%, can only be narrowed down by customizing the basis set for the Yb atom and/or the bond function for the Yb$_2$ molecule. The design principle of general-purpose bases may well be suboptimal for the particular system considered and the accuracy level required.} 

The X2C CCSD(T) potentials were used to approximate the BO Yb$_2$ potential represented by the model semianalytical function. On the one hand, semiclassical scaling to the known number of bound vibrational levels reduces the uncertainty in Yb$_2$ binding energy almost by an order of magnitude. This indicates that variations of the {\it ab initio} computational scheme alter the shape of the bound potential insignificantly. On the other, the constraint on the number of bound levels has its own uncertainty of the same order (30 cm$^{-1}$ or 4\%) as that of initial {\it ab initio} potential. It reflects the weak sensitivity of the ultracold PAS data to the potential minimum parameters. We found that the quoted uncertainty in $D_e$ is three orders of magnitude larger than the uncertainty in the position of the last bound vibrational level.

We also analyzed the effect of uncertainty in equilibrium distance established in the {\it ab initio} calculations. As a result, we bound the potential parameters of the BO Yb$_2$ potential as $R_e  \in [4.45, 4.55]$ {\AA} and $D_e \in [735, 775]$ cm$^{-1}$. This supports the reliability of the previous BO fit to PAS data  made with the potential having $R_e=4.52$ and optimized $D_e=743.0 \pm 2.4$  cm$^{-1}$ \cite{Borkowski2017}. We suggested modifying the model BO potential function to account for a strong correlation of the position and depth of the potential well.

Finally, we should stress that the sophisticated BO fit to PAS data \cite{Borkowski2017} was able to reduce uncertainty in the $D_e$ value from 20 to 2 cm$^{-1}$ to warranty $4\times 10^{-6}$ RMS deviation from the measured energies of the near-threshold levels. Thus, the new model function and its uncertainty obtained here promise remarkable improvements of above mentioned fits.

\begin{acknowledgments}
We thank Mateusz Borkowski, Piotr \.Zuchowski, and Dariusz K\k{e}dziera for many helpful discussions. Financial support by the Russian Science Foundation under project no. 17-13-01466 is gratefully acknowledged. 
P.T.~thanks an OPUS 17 research grant of the National Science Centre, Poland, (no.~2019/33/B/ST4/02114) and a scholarship for outstanding young scientists from the Ministry of Science and Higher Education.
\end{acknowledgments}

\end{document}